\documentclass[aps,prd,floatfix,reprint,twocolumn,reprint,amsmath,amssymb,superscriptaddress,showpacs]{revtex4-1}
\usepackage{graphicx}
\usepackage{bm}
\usepackage{amsmath}
\usepackage{dcolumn}
\usepackage{dsfont}
\usepackage{amssymb}
\usepackage{tabularx}
\usepackage{array}
\usepackage{float}
\usepackage{color}
\usepackage{epstopdf}
\usepackage{mathrsfs}
\usepackage[colorlinks, linkcolor=blue,anchorcolor=blue,citecolor=blue,urlcolor=blue]{hyperref}

\begin{document}
\title{Enhanced in-plane ferroelectricity, antiferroelectricity, and unconventional 2D emergent fermions in QL-XSbO$_2$ (X= Li, Na)}

\author{S. Guan}
\email{physguan@gmail.com}
\affiliation{State Key Laboratory of Superlattices and Microstructures, Institute of Semiconductors, Chinese Academy of Sciences, Beijing 100083, China}

\author{G. B. Zhang}
\affiliation{Institute for Computational Materials Science, School of Physics and Electronics, Henan University, Kaifeng  475004, China}
\affiliation{International Joint Research Laboratory of New Energy Materials and Devices of Henan Province, Kaifeng  475004, China}

\author{C. Liu}
\affiliation{Institute for Computational Materials Science, School of Physics and Electronics, Henan University, Kaifeng  475004, China}
\affiliation{International Joint Research Laboratory of New Energy Materials and Devices of Henan Province, Kaifeng  475004, China}

\begin{abstract}
Low-dimensional ferroelectricity and Dirac materials with protected band crossings are fascinating research subjects. Based on first-principles calculations, we predict the coexistence of spontaneous in-plane polarization and novel 2D emergent fermions in dynamically stable quadruple-layer (QL) XSbO$_2$ (X= Li, Na). Depending on the different polarization configurations, QL-XSbO$_2$ can exhibit unconventional inner-QL ferroelectricity and antiferroelectricity. Both ground states harbor robust ferroelectricity with enhanced spontaneous polarization of 0.56 nC/m and 0.39 nC/m for QL-LiSbO$_2$ and QL-NaSbO$_2$, respectively. Interestingly, the QL-LiSbO$_2$ possesses two other metastable ferroelectric (FE) phases, demonstrating the first 2D example with multiple FE orders. The ground FE phase can be flexibly driven into one of the two metastable FE phases and then into the antiferroelectric (AFE) phase. During this phase transition, several types of 2D fermions emerge, for instance, hourglass hybrid and type-II Weyl loops in the ground FE phase, type-II Weyl fermions in the metastable FE phase, and type-II Dirac fermions in the AFE phase. These 2D fermions are robust under spin-orbit coupling. Notably, two of these fermions, \textit{e.g.}, an hourglass hybrid or type-II Weyl loop, have not been observed before. Our findings identify QL-XSbO$_2$ as a unique platform for studying 2D ferroelectricity relating to 2D emergent fermions.
\end{abstract}

\maketitle
\section{Introduction}
Low-dimensional ferroelectrics have attracted significant interest recently for their importance in both fundamental science and functional electronics~\cite{Setter2006,Scott2007,Nuraje2013}. As conventional ferroelectric (FE) materials, such as perovskite oxides, are made thinner, the ferroelectricity tends to be suppressed owing to the presence of finite size effects~\cite{Batra1973,Zhong1994,Dawber2005,Wang2019d}. Two-dimensional (2D) materials exfoliated from their bulk counterparts preserve stable layered structures and may overcome the enhanced depolarization field, which should facilitate the appearance of ferroelectricity, thereby opening new perspectives for exploring low-dimensional ferroelectrics~\cite{Chang2016,Ding2017,Zhou2017,Hu2019,Guan2020}. So far, several atom-thick 2D materials have been proposed as ferroelectrics theoretically~\cite{Hu2019,Guan2020,He2019,Jia2019}. Particularly, some of them with hinge-like structures~\cite{Fei2016,Wan2017,Xiao2018,Guan2018,Zhang2018a,Liu2018b}, such as phosphorene structure, were predicted to possess in-plane FE orderings. However, none of these in-plane FE candidates have been verified in the experiment except for the ultrathin SnTe film~\cite{Chang2016,Liu2018d}, partially due to their low transition barriers~\cite{Guan2020,Hu2019}. Therefore, achieving a robust in-plane FE order is a critical bottleneck to the development of 2D ferroelectrics.

Meanwhile, inspired by graphene~\cite{CastroNeto2009}, much effort has been devoted to exploring topological materials with protected band crossing points, where the low-energy quasiparticles around these points can mimic the fermionic particles in high-energy physics~\cite{Volovik2003}, including Dirac~\cite{Elliott2015}, Weyl~\cite{Wan2011,Elliott2015}, and Majorana fermions~\cite{Armitage2018}. It was later discovered that more types of emergent fermions without high-energy analogs could also be realized in crystalline solids~\cite{Bradlyn2016}, such as type-II Weyl/Dirac~\cite{Soluyanov2015}, triple-point~\cite{Zhu2016}, nodal loop (NL)~\cite{Fang2016,Li2017}, and hourglass fermions~\cite{Wang2016}, which can display many unusual properties~\cite{Guan2017b,Yu2016,Sheng2017,Zhang2018}. Nevertheless, due to the reduced number of symmetries, the realization of unconventional quasiparticles in their 2D bulk counterparts, especially for symmetry-protected band crossings that are robust under spin-orbit coupling (SOC), is still a challenge. Young and Kane~\cite{Young2015} pointed out that nonsymmorphic operations can stabilize a truly Dirac point against SOC. Our previous work identified monolayer HfGeTe-family materials~\cite{Guan2017a} to host such 2D spin-orbit Dirac points. Recently, Wu \textit{et al.}~\cite{Wu2019} predicted monolayer GaTeI-family materials as a concrete material platform to realize 2D hourglass type-I Weyl NLs. However, the discovered 2D materials with novel emergent fermions are still limited. For example, a realistic 2D material with hourglass type-II or hybrid Weyl NLs has not been reported before.

From the above considerations, one see that it is desirable to explore 2D materials harboring various types of emergent fermions. Furthermore, for device applications, it is much more desirable to effectively tune these fermionic excitations utilizing static electric fields, owing to its compatibility with modern semiconductor technology. 2D ferroelectrics with puckered hinge structures and multiple polarization orders naturally serve as good candidates to realize these objectives: they typically feature certain nonsymmorphic symmetries, leading to the emergence of symmetry-protected 2D fermions that are robust under SOC; moreover, the electric-field-enforced transition between different polarization orders allows the electric control of the unconventional fermionic excitations.

In this work, using first-principles calculations, we demonstrate the presence of in-plane FE and antiferroelectric (AFE) orders together with diverse types of 2D emergent fermions in quadruple-layer (QL) XSbO$_2$ (X= Li, Na), which consists of two middle XO layers and top-and-bottom SbO layers. Each atomic layer possesses a hinge-like structure with O atoms slightly displaced from other atoms along the armchair direction, resulting in the emergence of enhanced in-plane spontaneous polarization. Interestingly, two metastable FE phases appear for QL-LiSbO$_2$, endowing itself with multiple FE orders. More importantly, using QL-LiSbO$_2$ as a prototype, it demonstrates 2D hourglass hybrid and type-II Weyl NLs coexisting in the Brillouin zone (BZ) for the ground FE phase. An applied external electric field can drive the ground FE phase into one metastable FE phase and then into an AFE phase. Correspondingly, the hourglass type-II Weyl NLs can be transformed into type-II Weyl fermions and type-II Dirac fermions for the metastable FE phase and the AFE phase, respectively. All types of 2D emergent fermions discovered here are symmetry-protected against SOC. Our study discovers good candidates for realizing robust in-plane ferroelectricity; additionally, it offers a promising material platform for exploring the interplay between ferroelectricity, antiferroelectricity, and unconventional 2D emergent fermions.
 
\section{Computational details}
Our first-principle calculations were carried out by employing Vienna \textit{ab-initio} simulation package (VASP)~\cite{Kresse1993,Kresse1996} within the framework of the projector augmented wave (PAW) method~\cite{Bloechl1994}. The Perdew-Burke-Ernzerhof (PBE)~\cite{PBE} implementation of the generalized gradient approximation (GGA) was adopted for the exchange-correlation potential. The structures were fully relaxed until the energy and force converged to less than 10$^{-5}$ eV and 10$^{-2}$ eV/\AA, respectively. The plane-wave energy cutoff was set to be 520 eV, and the BZ integration was performed on Monkhorst-Pack $k$ mesh of size 12$\times$14$\times$1 for QL-LiSbO$_2$ and 10$\times$12$\times$1 for QL-NaSbO$_2$, respectively. The optimized van der Waals (vdW) correlation functional optB86b-vdW \cite{Klimefmmodeheckslsesi2011} has been taken into account in the calculation of exfoliation energy. A vacuum layer with thickness larger than 15 \AA \ was placed to avoid artificial interactions between periodic images. The phonon spectrum was obtained by using PHONOPY code through the DFPT method~\cite{Togo2015}. The ferroelectric polarization was calculated using the Berry phase method~\cite{King-Smith1993,Resta1994}. The minimum energy pathways of polarization reversal transitions are obtained through the climbing image nudged elastic band (CINEB) method~\cite{Henkelman2000}.

\section{Results}

\subsection{Structure and stability}

\begin{figure}[!t]
\centerline{\includegraphics[width=0.49\textwidth]{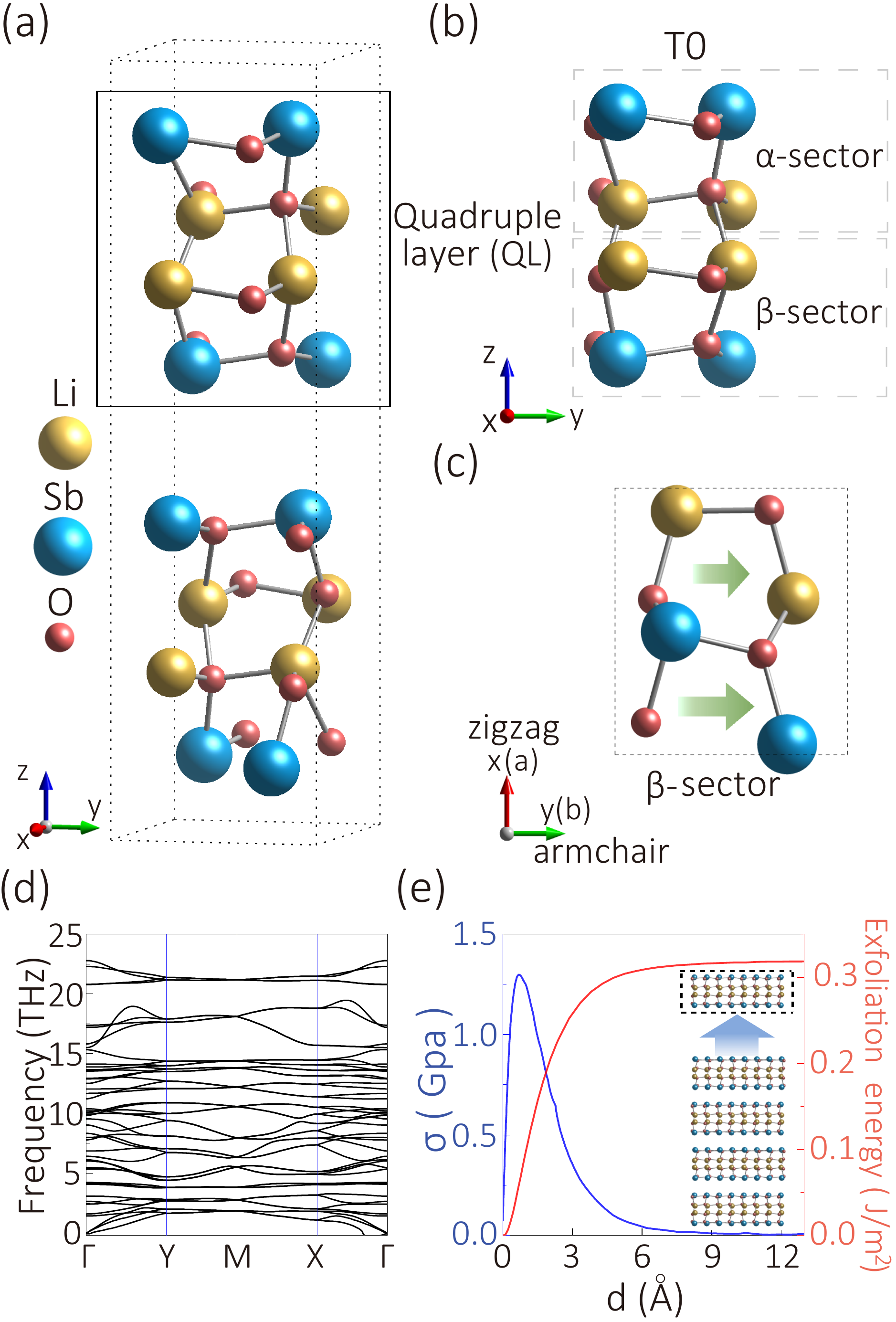}}
\caption{(a) Crystal structure of layered bulk LiSbO$_2$. The black square indicates a quadruple layer (QL). (b) Side view of QL-LiSbO$_2$, which has been divided into $\alpha$- and $\beta$-sectors. (c) Bottom view of QL-LiSbO$_2$ with only the $\beta$-sector plotted. The dashed square shows the unit cell. The green arrows indicate the small dipole moments in each atomic layer. (d) Phonon spectrum of QL-LiSbO$_2$, showing the dynamic stability of the structure. (e) Calculated exfoliation energy (red line) for QL-LiSbO$_2$ as a function of the distance separated from the bulk (as shown in the inset). The blue curve represents the exfoliation strength $\sigma$.}
\label{fig1}
\end{figure} 

Both bulk LiSbO$_2$ and NaSbO$_2$ are ternary antimony oxides with similar layered structures~\cite{Stover1980,McColm1994}. As a representative, three-dimensional (3D) LiSbO$_2$ is shown in Fig.~\ref{fig1}(a). Four covalently bonded atomic layers are grouped into an SbO-LiO-LiO-SbO quadruple layer (QL), and the QLs are weakly bonded via vdW interactions, stacking along the crystal c axis, that is, the $z$ direction. In its 2D form, QL-LiSbO$_2$ has a nonsymmorphic $Pca2_1$ space group with a buckled conformation (denoted as the T0 phase). The optimized lattice parameters $a$ and $b$ for T0 QL-LiSbO$_2$ are 5.694 and 4.974 \AA, respectively. We further calculate its phonon spectrum and confirm that the structure is dynamically stable due to the absence of any imaginary frequency across the BZ (see Fig.~\ref{fig1}(d)). 

Fig.~\ref{fig1}(b) shows the side view of T0 QL-LiSbO$_2$, in which the four atomic layers are divided into two sectors (labeled as $\alpha$- and $\beta$-sectors). Each sector comprises two atomic chains and the links within the chains provide a zigzag arrangement along the $x$ direction, showing a puckered structure similar to SnTe monolayer~\cite{Chang2016,Liu2018d}, as illustrated by the $\beta$-sector in Fig.~\ref{fig1}(c). Notably, in each sector, both the Li and Sb atoms have a small displacement along the armchair ($y$) direction with respect to the O atoms, forming dipole moments in each atomic layer. These small dipole moments strongly hint at the emergence of in-plane spontaneous electric polarization. More importantly, since the two sectors are related to each other via a glide mirror operation, the dipoles in different sectors must have the same orientation, doubling the total dipole moments for the whole T0 QL-LiSbO$_2$. Hence, enhanced polarization in the material is expected.

To check the possibility of obtaining T0 QL-LiSbO$_2$ from its bulk samples via mechanical exfoliation, we evaluate the cleavage energy and exfoliation strength in Fig.~\ref{fig1}(e). With increasing separation distance ($d$) between the simulated QL and the bulk sample, the energy initially increases quickly and then saturates to a value corresponding to the exfoliation energy of approximately 0.32 J/m$^2$. This value, which is slightly less than that of graphene (0.37 J/m$^2$)~\cite{Zacharia2004} and MoS$_2$ (0.41 J/m$^2$)~\cite{Bjoerkman2012}, is in the typical range for the vdW-layered compounds~\cite{Bjoerkman2012}. The exfoliation strength, defined as the maximum derivative of the exfoliation energy with respect to the separation distance $d$, is approximately 1.31 GPa, which is also less than that of graphene (2.10 GPa). In addition, similar values (0.31 J/m$^2$ and 1.30 GPa) were obtained for QL-NaSbO$_2$. Our results thus suggest that QL-XSbO$_2$ (X= Li, Na) can be exfoliated from its bulk counterparts.

\begin{figure}[!htb]
\centerline{\includegraphics[width=0.49\textwidth]{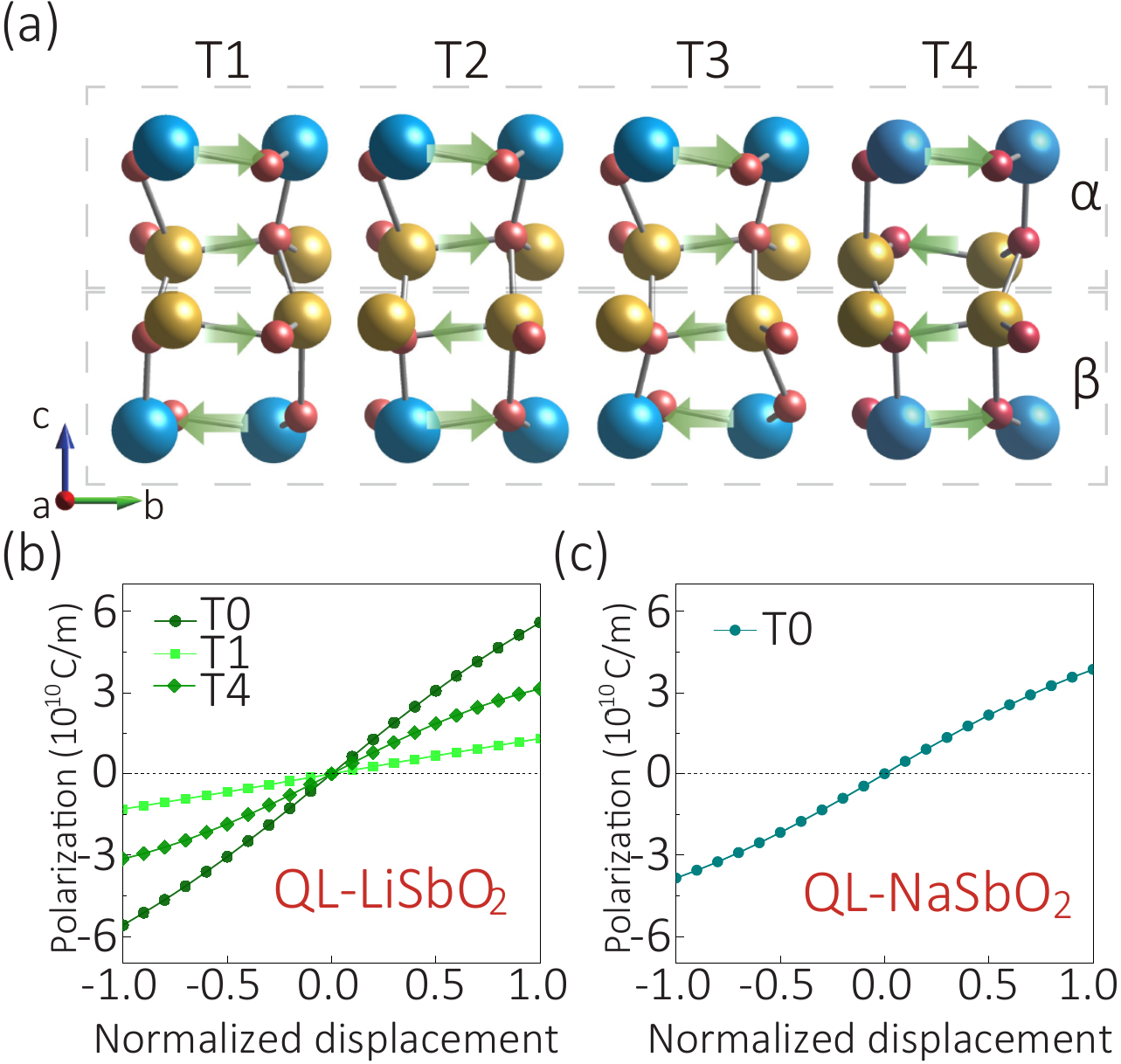}}
\caption{(a) Side views of T1, T2, T3 and T4 QL-LiSbO$_2$ with corresponding polarization configurations. The green arrows indicate the polarization arrangement. Calculated spontaneous polarization of (b) QL-LiSbO$_2$ and (c) QL-NaSbO$_2$ as a function of the normalized displacement along the adiabatic path.}
\label{fig2}
\end{figure}

\subsection{Ferroelectricity and antiferroelectricity}

\begin{table*}[!htb]
\caption{The basic properties of QL-XSbO$_2$ (X= Li, Na) are listed. $a$ and $b$ are the lattice constants, $P_s$ is the spontaneous polarization, $E_{tot}$ is the total energy per unit cell with respect to the T0 phase, and $E_{gap}$ is the indirect band gap. We also list the symmetry operations except for the identity operation. $\mathcal{P}$ denotes the inversion symmetry. }
\label{tab:tab1}
\begin{tabular}{p{1.8cm}<{\centering}|p{2.2cm}<{\centering}|p{1cm}<{\centering}| p{1cm}<{\centering}| p{2.1cm}<{\centering}| p{1.8cm}<{\centering}| p{1.8cm}<{\centering}| p{4.5cm}<{\centering}}
  \hline\hline
   QL-LiSbO$_{2}$ & space group &  $a$({\AA}) & $b$({\AA}) & $P_s$(10$^{-10}$C/m) & $E_{tot}$(meV)  & $E_{gap}$(eV)  & symmetry \\ \hline
    T0 & Pca2$_1$ & 5.694 & 4.974 & 5.586 &  0  & 3.176  & $\{M_z|\frac{1}{2}0$\}, $\{M_x|\frac{1}{2}\frac{1}{2}$\}, $\{C_y|0\frac{1}{2}$\} \\ \hline
    T1 & Pc & 5.692 & 4.980 & 1.300 &  124.1  & 3.158  & $\{M_x|\frac{1}{2}\frac{1}{2}$\} \\ \hline
    T3 & P2$_1$/c & 5.723 & 5.052 & AFE &  63.3  & 3.229  & $\mathcal{P}$, $\{M_x|\frac{1}{2}\frac{1}{2}$\}, $\{C_x|\frac{1}{2}\frac{1}{2}$\} \\ \hline 
    T4 & Pca2$_1$ & 5.660 & 5.117 & 3.141 &  230.2  & 3.304  & $\{M_z|\frac{1}{2}0$\}, $\{M_x|\frac{1}{2}\frac{1}{2}$\}, $\{C_y|0\frac{1}{2}$\} \\ \hline \hline
   QL-NaSbO$_{2}$ & space group  & $a$({\AA}) & $b$({\AA}) & $P_s$(10$^{-10}$C/m) & $E_{tot}$(meV)  & $E_{gap}$(eV)  & symmetry \\ \hline
    T0 & Pca2$_1$ & 6.435 & 4.922 & 3.854 &  0  & 3.428 & $\{M_z|\frac{1}{2}0$\}, $\{M_x|\frac{1}{2}\frac{1}{2}$\}, $\{C_y|0\frac{1}{2}$\} \\ \hline
    T3 & P2$_1$/c & 6.433 & 4.914 & AFE &  1.6  & 3.421  & $\mathcal{P}$, $\{M_x|\frac{1}{2}\frac{1}{2}$\}, $\{C_x|\frac{1}{2}\frac{1}{2}$\}\\
  \hline\hline
\end{tabular}
\end{table*} 
\begin{figure*}[!t]
\centerline{\includegraphics[width=0.85\textwidth]{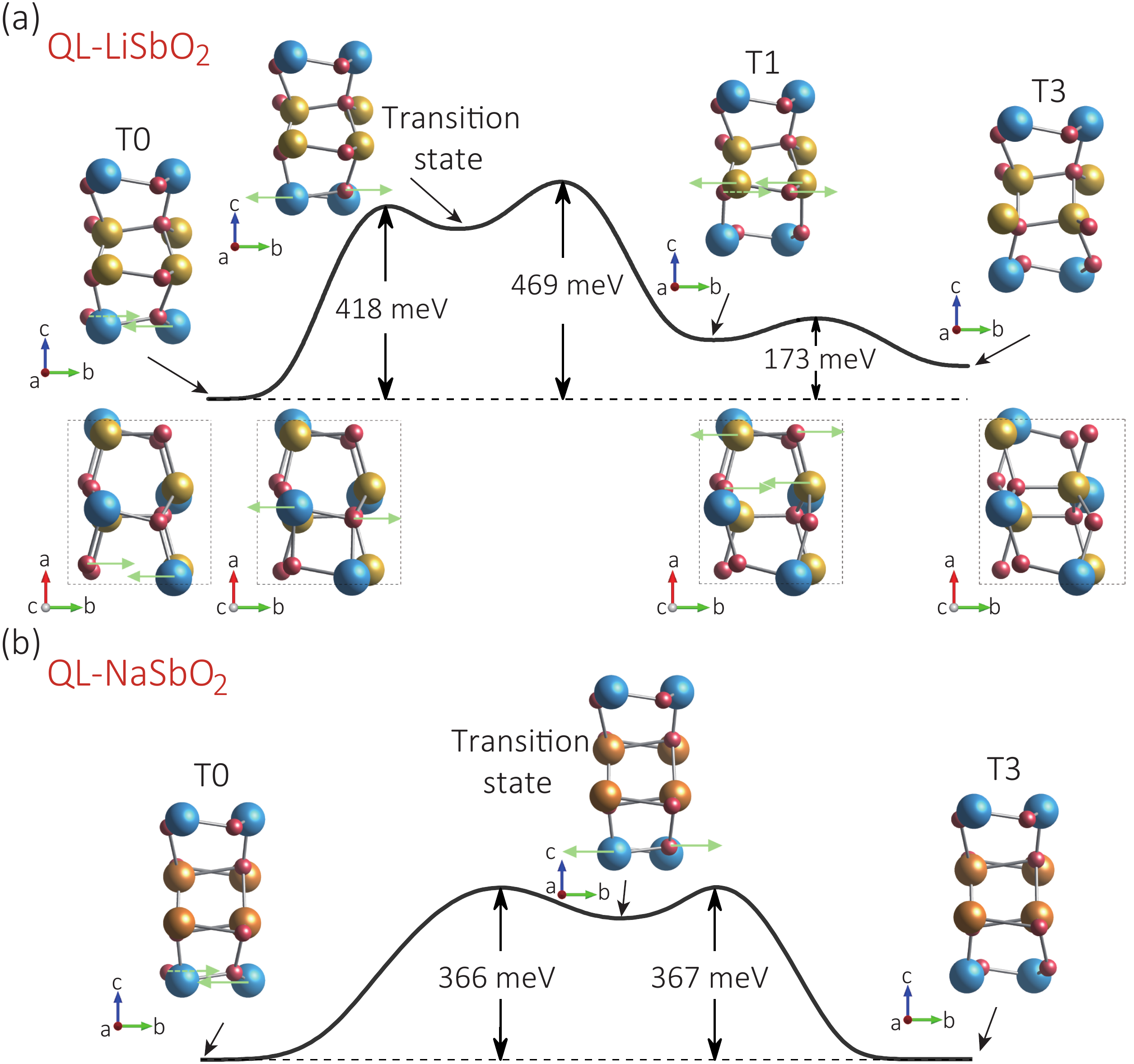}}
\caption{Kinetic pathways of polarization reversal processes. (a) Energy profile of the most effective kinetic pathway of one QL-LiSbO$_2$ transforming from the ground FE T0 phase to the AFE T3 phase, which involves a three-step concerted mechanism. (b) Energy profile of the transition from the T0 phase to the T3 phase for QL-NaSbO$_2$, during which a transition state is undergone. The green arrows attached to atoms refer to the directions of atomic motion during the polarization reversal processes. Note that only half of the polarization reversal process for T0 QL-XSbO$_2$ is plotted for simplicity (the other half can be considered as an inverse process and the rest of the energy profile can be symmetrically obtained).}
\label{fig3}
\end{figure*}
The enhanced spontaneous polarization in exfoliated T0 QL-XSbO$_2$ was confirmed by our DFT calculations. Using the modern Berry phase method~\cite{King-Smith1993,Resta1994}, we reveal that the T0 QL-XSbO$_2$ phases have significant polarization along the $y$ direction, with a sizable magnitude of up to 0.56 nC/m and 0.39 nC/m for QL-LiSbO$_2$ and QL-NaSbO$_2$, respectively (see Figs.~\ref{fig2}(b-c)). The calculated in-plane polarization ($P_s$) is much larger than the value for monolayer SnTe (∼0.19 nC/m)~\cite{Wan2017} that was successfully detected in the experiment~\cite{Chang2016}. This is attributed to the unique stacking structure mentioned above, which enables the T0 QL-XSbO$_2$ to feature a distinctive inner-QL polarization configuration with the same orientation in each atomic layer.

Since the spontaneous polarization in each atomic layer holds the potential to be reversed, we systematically examine other possible structures of atomic configuration within a QL, which may emerge during the process of polarization switching and display different FE or AFE orders. Fig.~\ref{fig2}(a) illustrates the candidate structures with different polarization arrangements, including the polarization reversal of the $\beta$-SbO layer (denoted as T1 phase), the polarization reversal of the $\beta$-XO layer (denoted as T2 phase), the polarization reversal of both the $\beta$-SbO and $\beta$-XO layers (denoted as T3 phase), and the polarization reversal of both the $\alpha$- and $\beta$-XO layers (denoted as T4 phase). We further calculated the phonon spectra and identified the dynamic stability of the T1, T3, and T4 phases for QL-LiSbO$_2$ and the T3 phase for QL-NaSbO$_2$, respectively (Fig. S1,S2).

The optimized lattice parameters and total energy of the dynamically stable structures are summarized in Table~\ref{tab:tab1}. For QL-LiSbO$_2$ and QL-NaSbO$_2$, we find that both the ground states are FE T0 phase, and they also possess an AFE T3 phase with the opposite polarization in each sector. For QL-LiSbO$_2$, the total energy of the T3 phase is 63.3 meV higher than that of the T0 phase, whereas the two phases are nearly degenerate with a slight energy difference of 1.6 meV for QL-NaSbO$_2$. Surprisingly, there are two other FE phases for QL-LiSbO$_2$, which are metastable with much higher energies around 124 meV (T1 phase) and 230 meV (T4 phase). The calculated polarization for the metastable FE phases is 0.13 nC/m and 0.31 nC/m for the T1 and T4 phases, respectively (see Fig.~\ref{fig2}(b)), demonstrating QL-LiSbO$_2$ as an unconventional inner-QL ferroelectrics with multiple FE orders.

To inspect the robustness of the ferroelectricity and the feasibility of polarization switching, we next investigate the most effective kinetic pathway connecting the two degenerate states with opposite polarities for T0 QL-XSbO$_2$. First, we consider the situation in which the lattice constants are fixed to the relaxed values of the FE T0 phase during the polarization reversal process. We obtain that the minimal energy barrier to reverse the polarization in both $\alpha$- and $\beta$-sectors of T0 QL-XSbO$_2$ is to reverse one of them first and then the other. Therefore, the AFE T3 phase is undergone in the process (see Fig. S4 for more possible kinetic pathway results). 

In Fig.~\ref{fig3}(a), a three-step concerted motion of the lower two atomic layers in the $\beta$-sector is revealed in the transition from the T0 phase to the T3 phase for QL-LiSbO$_2$. In the first step, the dipole between the Sb and O atoms of the $\beta$-SbO layer along the armchair direction tends to weaken until it reaches zero, forming a transition structure. In the second step, the transition structure transforms into the metastable T1 phase by gradually increasing the dipole moment in the opposite direction, leading to the polarization reversal of the $\beta$-SbO layer. In the third step, the dipole moment in the $\beta$-LiO layer is reversed, finally reaching the AFE T3 phase. In contrast, a similar but simplified two-step process was found for T0 QL-NaSbO$_2$, during which the transition state has no polarization in the $\beta$-SbO layer (see Fig.~\ref{fig3}(b)). The overall activation barriers of the concerted processes are 469 meV and 367 meV per unit cell, which correspond to 117 meV/f.u. and 92 meV/f.u for QL-LiSbO$_2$ and QL-NaSbO$_2$, respectively. The obtained values are much larger than that of monolayer SnTe (∼4.5 meV/f.u.)~\cite{Wan2017}, and are comparable to those of 2D In$_2$Se$_3$ (∼69 meV/f.u.)~\cite{Ding2017} and conventional ferroelectric PbTiO$_3$ (∼200 meV/f.u.)~\cite{Cohen1992}, indicating their robust ferroelectricity. We also performed \textit{ab-initio} molecular dynamics simulations, and the results show that their ferroelectricity can be maintained above room temperature (Fig. S3).

Next, we consider the situation in which all the structural parameters are fully relaxed during the polarization reversal process for T0 QL-XSbO$_2$. We find that the overall activation barriers are significantly reduced to 31 meV/f.u. and 52 meV/f.u. for QL-LiSbO$_2$ and QL-NaSbO$_2$, respectively (Fig. S5). In particular, the reversal process for QL-LiSbO$_2$ now undergoes a transition structure that corresponds to a paraelectric (PE) phase where the dipole moments in each layer are zero, whereas the case for QL-NaSbO$_2$ preserves its main process. Hence, our results demonstrate that the energy barriers are flexibly tunable, providing QL-XSbO$_2$ with a wider application prospect.

\subsection{2D Hourglass Weyl loops}
\begin{figure}[!b]
\centerline{\includegraphics[width=0.49\textwidth]{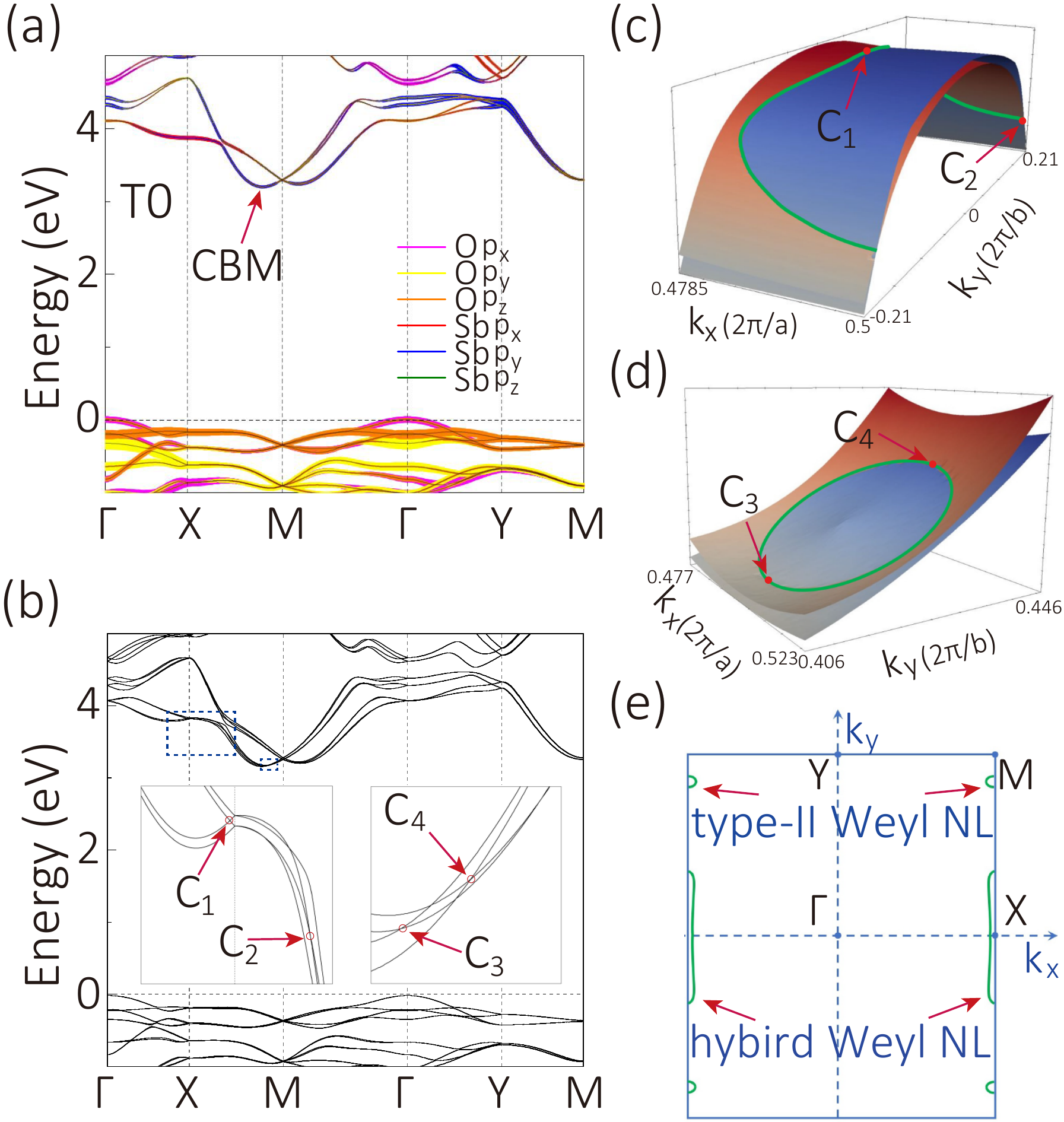}}
\caption{Band structure of T0 QL-LiSbO$_2$ (a) without SOC and (b) with SOC. Orbital-contributions in panel (a) are represented by color lines with thickness proportional to the orbital weight. Insets in panel (b) plot the band structures zoomed around the crossing points marked by the blue dashed squares. 2D band structures around the hourglass (c) hybird Weyl NL and (d) type-II Weyl NL, which are indicated by green color. (e) Shape of hourglass Weyl NLs obtained from DFT calculations.}
\label{fig4}
\end{figure}
Now we turn to study the electronic structure of QL-XSbO$_2$. Since the two materials show qualitatively similar behavior, our following discussion will be mainly based on the results of QL-LiSbO$_2$ (see Fig. S6 for the results of QL-NaSbO$_2$). We first consider the ground T0 phase, in which the symmetry operations involve the glide mirror \{$M_z|\frac{1}{2}0$\}, the glide mirror \{$M_x|\frac{1}{2}\frac{1}{2}$\}, the twofold screw rotation \{$C_y|0\frac{1}{2}$\}, and the time reversal symmetry $\mathcal{T}$. Fig.~\ref{fig4}(a) shows its calculated band structure without SOC. One observes that the conduction bands are mostly composed of the hybridized Sb-5p and O-2p orbitals, whereas the valence bands are dominated by the O-2p orbitals with a small admixture of the pz state of the Sb atoms. In addition, the T0 QL-LiSbO$_2$ is an indirect band-gap semiconductor with the valence band maximum located at the $\Gamma$ point and the conduction band minimum (CBM) located along the X-M symmetry line. The band gap is calculated to be approximately 3.1 eV.

As can be expected, including SOC strongly affects the electronic structure of the conduction bands (see Fig.~\ref{fig4}(b)). Owing to the absence of inversion symmetry broken by the spontaneous polarization, SOC produces the general band spin splittings. Interestingly, hourglass-type dispersions emerge on the $\Gamma$-X and X-M paths with several band crossing points marked by C1 around the X point, and C2, C3, and C4 along the X-M path, as shown in the insets of Fig.~\ref{fig4}(b). Unlike the C1 point featuring a normal linear crossing, the two linear crossing bands around the C2, C3, and C4 points are tilted with the same sign of slopes along the X-M (k$_y$) direction, demonstrating a type-II crossing behavior. 

In fact, due to the presence of the glide mirror $\widetilde{\mathcal{M}}_z=\{M_z|\frac{1}{2}0$\} in the T0 phase, the neck crossing points of the hourglass dispersions should trace out two different types of hourglass Weyl NLs: one is a hybrid Weyl NL surrounding the X point (C1 and C2), and the other belongs to a type-II Weyl NL on the X-M path (C3 and C4). We scan the BZ and plot the 2D dispersion of these NLs in Figs.~\ref{fig4}(c-d). The calculation indeed confirms the existence of two distinctive Weyl NLs in the band structure. Note that the loops here survive under SOC, which are different from those of monolayer Cu$_2$Si~\cite{Feng2017} and CuSe~\cite{Gao2018} that have been discovered in the experiment. Moreover, they also differ from the 2D hourglass type-I Weyl NLs predicted in monolayer GaTeI~\cite{Wu2019} and surface-supported Bi/Cl-SiC(111)~\cite{Wang2019c} for their unique band dispersion. 

The mechanism that protects the Weyl NLs against SOC can be clarified by the following symmetry analysis. Because any $k$ point in the BZ is invariant under $\widetilde{\mathcal{M}}_z$, each Bloch state $|u\rangle$ has a definite eigenvalue of $\widetilde{\mathcal{M}}_z$, which is given by $g_z=\pm ie^{-ik_x/2}$. The eigenvalues depend only on $k_x$; consequently, at the time reversal invariant point X ($k_x=\pi$), we have $g_z=\pm 1$. As a result, the degenerate Kramers pairs $|u\rangle$ and $\mathcal{T}|u\rangle$ share the same $\widetilde{\mathcal{M}}_z$ eigenvalue. Nevertheless, at the $\Gamma$ point ($k_x=0$), owing to $g_z=\pm i$, each Kramers pair must possess the opposite $g_z$. Such a different pairing at X and $\Gamma$ guarantees the hourglass dispersion of the four bands~\cite{Wang2016}, which stems from the switch of partners between two pairs when going from X to $\Gamma$ along one given path~\cite{Guan2017b,Wu2019}. Because the path can be arbitrary and there must exist at least one Weyl crossing point on the path, the resultant hourglass Weyl NLs formed by these crossing points emerge in the BZ, as shown by their shapes in Fig.~\ref{fig4}(e). Therefore, both the Weyl NLs in T0 QL-LiSbO$_2$ are symmetry-protected with the two crossing bands having opposite $g_z$.

\subsection{2D type-II spin-orbit Weyl/Dirac points}
\begin{figure}[!htb]
\centerline{\includegraphics[width=0.5\textwidth]{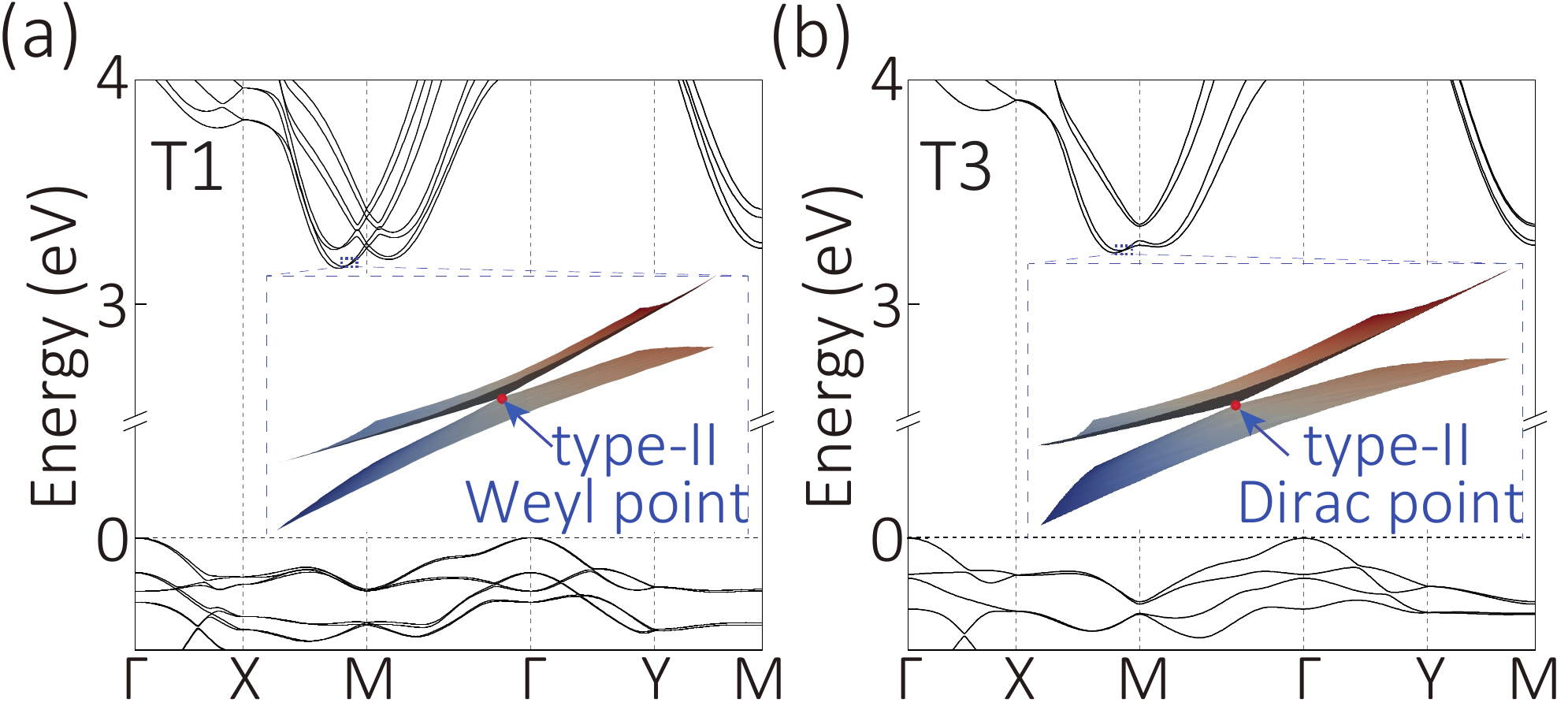}}
\caption{Band structures of (a) T1 and (b) T3 QL-LiSbO$_2$ with SOC included. Insets are the corresponding 2D plots of the band structure around the crossing point indicated by the dashed box, which shows a type-II Weyl point and a type-II Dirac point for the T1 and T3 phases, respectively.}
\label{fig5}
\end{figure}
The glide mirror $\widetilde{\mathcal{M}}_z$ is broken once the T0 QL-LiSbO$_2$ is driven into the metastable T1 and T3 phases, opening gaps at the original hourglass Weyl NLs. However, due to the presence of other nonsymmorphic operations (see Table~\ref{tab:tab1}), one may wonder: Is it possible to have new types of 2D emergent fermion states that are also against SOC? To address this issue, we first focus on the T1 phase, which possesses only one nonsymmorphic operation, $\widetilde{\mathcal{M}}_x=\{M_x|\frac{1}{2}\frac{1}{2}$\}. Fig.~\ref{fig5}(a) shows the calculated band structure including the SOC. Interestingly, one observes that there emerges a linear crossing point around the CBM. The zoomed-in plot further validates that it is an isolated Weyl point. In addition, this Weyl cone is tipped over, indicating a 2D type-II spin-orbit Weyl point.

We find that this type-II Weyl point is also symmetry-protected. The reason is that the path X-M at $k_x=\pi$ is an invariant subspace for $\widetilde{\mathcal{M}}_x$, so each Bloch state $|u\rangle$ on this path has a definite eigenvalue of $\widetilde{\mathcal{M}}_x$, which is given by $g_x=\pm ie^{-ik_y/2}$. Consequently, the crossing point formed by two bands would be a protected Weyl point if the two bands have opposite $g_x$ on the X-M path. This is exactly the case for the type-II Weyl point here. For a generic $k$ point deviating from the X-M path, it will no longer be protected by $\widetilde{\mathcal{M}}_x$, thereby generating an isolated Weyl point on X-M.

In the following, we study the AFE T3 phase, where the inversion symmetry $\mathcal{P}$ is present. Therefore, each band here is doubly degenerate due to the combined $\mathcal{PT}$ symmetry (see Fig.~\ref{fig5}(b)). A linear crossing point with a type-II behavior on the X-M path is seen again, which is located close to the CBM but now is fourfold degenerate. Such a crossing point suggests a 2D type-II spin-orbit Dirac point, as confirmed in the inset of Fig.~\ref{fig5}(b). Furthermore, since the $\widetilde{\mathcal{M}}_x$ is preserved, the commutation relation between $\widetilde{\mathcal{M}}_x$ and $\mathcal{PT}$ on X-M takes the form of $\widetilde{\mathcal{M}}_x\mathcal{PT}=T_{11}\mathcal{PT}\widetilde{\mathcal{M}}_x=-e^{-ik_y}\mathcal{PT}\widetilde{\mathcal{M}}_x$. Accordingly, the Kramers degenerate pair $|u\rangle$ and $\mathcal{PT}|u\rangle$ have the same $\widetilde{\mathcal{M}}_x$ eigenvalue $g_x$, because we have $\widetilde{\mathcal{M}}_x(\mathcal{PT}|u\rangle)=g_x(\mathcal{PT}|u\rangle)$. Therefore, similar to the type-II Weyl point in the T1 phase, the type-II Dirac point with fourfold degeneracy on X-M is protected for the opposite $g_x$ hosted by the two crossing bands.

\section{Discussion and conclusion}
We discuss a few points before closing. We emphasize that T0 QL-XSbO$_2$ possesses enhanced in-plane spontaneous polarization and high transition barrier simultaneously, which are keys to the development of 2D ferroelectrics. Meanwhile, the multiple polarization orders in QL-XSbO$_2$ are rare in 2D materials, which can exhibit unique polarization reversal and structural phase-transition features. Such distinctive features provide a route towards many device applications, such as 2D dielectric capacitors~\cite{Jia2019}, and barrier layers in tunnel junctions~\cite{Apachitei2017}. 

We have revealed QL-XSbO$_2$ as a fertile playground to study hourglass Weyl NL, and type-II spin-orbit Weyl/Dirac fermions in 2D. Moreover, we report two novel types of fermions, e.g., 2D hourglass hybrid and type-II Weyl NLs. Many of their physical properties may be different from those of type-I Weyl NLs, such as suppressed optical absorption and unusual magneto-oscillations~\cite{Yu2016,Zhang2018}. In addition, we envision a possibility of observing interesting valleytronic physics similar to graphene and 2D transition metal dichalcogenides~\cite{Xiao2012,Mak2014}, due to the presence of two valleys in the conduction band. In particular, the polarization switching between FE and AFE phases may also enable tunable valleytronics via static electric field~\cite{Yu2020}.

Although QL-LiSbO$_2$ and QL-NaSbO$_2$ are wide band gap semiconductors, the 2D emergent fermions found here are close to the CBM; hence, a moderate electron doping is sufficient to observe their properties. It is known that oxygen vacancies naturally form in metal oxides that produce n-type conductivity~\cite{Bierwagen2015}, thereby facilitating the further experimental study. In contrast, searching for more ideal materials with electrically controlled 2D emergent fermions located precisely at the Fermi level will be an appealing future research. Guided by the findings in this work, we point out that 2D ferroelectric metals~\cite{Shi2013,Zhou2020} with certain nonsymmorphic operations are good candidates for achieving this goal.

In conclusion, we discovered the intrinsic in-plane ferroelectricity and antiferroelectricity in QL-XSbO$_2$ (X= Li, Na), which also hold a variety of 2D emergent fermions that are tunable by polarization switching. Our first-principles calculation shows that the materials are stable, which can be readily obtained from their 3D bulk counterparts. Both ground states are FE with large spontaneous polarization, and their ferroelectricity is robust owing to the high transition barriers. We show that all types of emergent fermions are symmetry-protected against SOC, including the 2D hourglass hybrid and type-II Weyl loops in the ground FE phase, the 2D type-II Weyl fermions in the metastable FE phase, and the 2D type-II Dirac fermions in the AFE phase. Our results thus provide an excellent platform for exploring the intriguing physics of 2D ferroelectrics associated with electric-controlled 2D emergent fermions, leading to potential applications in nanoscale devices.

\begin{acknowledgments}
The work is supported by the National Key R\&D Program of China (Grant No. 2018YFB2202802), the NSF of China (Grant No.11904359), and the Strategic Priority Research Program of Chinese Academy of Sciences (Grant No.XDB30000000). C. L. thanks the funding support from the National Natural Science Foundation of China (Grant No.11904079) and the China Postdoctoral Science Foundation (No.2019M652303).
\end{acknowledgments}

%


\end{document}